# [(Li$_{0.8}$Fe$_{0.2}$)OH]FeS and the ferromagnetic superconductors [(Li$_{0.8}$Fe$_{0.2}$)OH]Fe(S$_{1-x}$Se$_x$) (0 < x < 1)

Ursula Pachmayr and Dirk Johrendt*

Department Chemie, Ludwig-Maximilians-Universität München
Butenandtstr. 9-13, 81377 München, Germany, E-mail: johrendt@lmu.de

**Superconductivity up to 43 K and ferromagnetic ordering coexist in the iron chalcogenides [(Li$_{0.8}$Fe$_{0.2}$)OH]Fe(S$_{1-x}$Se$_x$) (0 < x ≤ 1). Substitution of sulphur for selenium gradually supresses superconductivity while ferromagnetism persists up to non-superconducting [(Li$_{0.8}$Fe$_{0.2}$)OH]FeS.**

A conclusive understanding of unconventional superconductivity in correlated electron systems is among the most challenging topics in contemporary solid state chemistry and physics.[1] In copper-oxide[2] and iron-based[3] materials, superconductivity emerges close to the disappearance of an antiferromagnetically ordered ground state,[4, 5] leading to the assumption that magnetic fluctuations play a crucial role in the formation of the cooper pairs.[6] In contrast, superconductivity is generally considered to be incompatible with ferromagnetism. The latter generates magnetic flux, while superconductivity expels magnetic flux from the interior of a solid. Nevertheless, a few examples where both forms of order coexist are known (see Refs. 1-14 in [7]). However, a detailed examination of the fascinating coexistence phenomena is mostly aggravated by extremely low transition temperatures, as well as the inertness of the rare-earth 4$f$ shell with respect to the chemical environment.

Recently we reported the ferromagnetic iron selenide superconductor [(Li$_{1-x}$Fe$_x$)OH](Fe$_{1-y}$Li$_y$)Se.[7] The crystal structure exhibits alternately stacked lithium-iron-hydroxide layers and iron selenide layers, and was contemporaneously observed by *Lu* et al.[8] and by *Sun* et al..[9] Electron doping of the FeSe layer is most probably the main reason for the enormous increase of $T_c$ from 8 K in β-FeSe[10] to 43 K in [(Li$_{1-x}$Fe$_x$)OH](Fe$_{1-y}$Li$_y$)Se. Similar effects on $T_c$ were found molecular intercalated iron selenides like Li$_x$(NH$_2$)$_y$(NH$_3$)$_{1-y}$Fe$_2$Se$_2$[11] or Li$_x$(C$_5$H$_5$N)$_y$F$_{2-z}$Se$_2$.[12] However, the coexistence of unconventional superconductivity and ferromagnetism in [(Li$_{1-x}$Fe$_x$)OH](Fe$_{1-y}$Li$_y$)Se is striking. Even though the internal dipole field of the ferromagnet acts on the superconductor, superconductivity is not suppressed whereby ferromagnetism and superconductivity coexist in a spontaneous vortex phase. Influencing one of these order parameters would give the opportunity to examine the fascinating interplay i.e. the competition, coexistence and coupling of ferromagnetism and superconductivity in more detail.

In this communication we present the new chalcogenides [(Li$_{0.8}$Fe$_{0.2}$)OH]Fe(S$_{1-x}$Se$_x$) (0 ≤ x < 1). We show that the gradual substitution of selenium by sulphur continuously reduces the critical temperature until superconductivity is absent in the pure sulphide. This allows studying the influence of chemical pressure, and in particular possible effects on the coexistence of superconductivity and ferromagnetism.

Polycrystalline samples of [(Li$_{0.8}$Fe$_{0.2}$)OH]Fe(S$_{1-x}$Se$_x$) were synthesized under hydrothermal conditions.[7] Iron metal (0.0851 g), LiOH·H$_2$O (3 g) and appropriate amounts of thiourea respectively selenourea were mixed with distilled water (10 mL). The starting mixtures were tightly sealed in a teflon-lined steel autoclave (50 mL) and heated at 155 °C for 7 days. After washing with distilled water and ethanol, the polycrystalline products were dried at room temperature under dynamic vacuum and stored at −25 °C under argon atmosphere. Structural characterization by X-ray powder diffraction (PXRD) revealed single phase samples of [(Li$_{0.8}$Fe$_{0.2}$)OH]Fe(S$_{1-x}$Se$_x$) which is isostructural to the selenide.[7-9] Figure 1 shows the Rietveld-fit of [(Li$_{0.8}$Fe$_{0.2}$)OH]FeS.

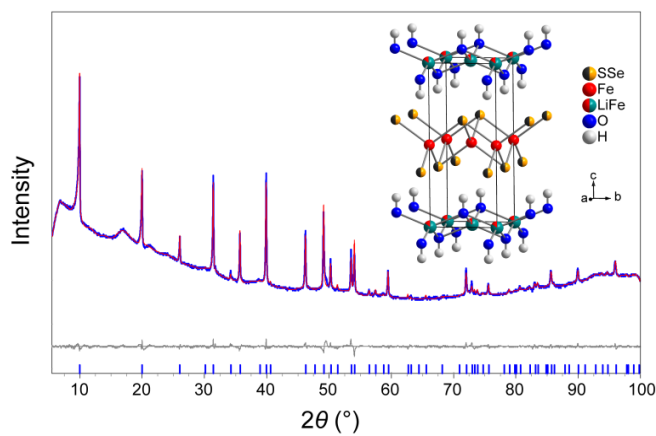

**Fig. 1** X-ray powder pattern of [(Li$_{0.8}$Fe$_{0.2}$)OH]FeS (blue) with Rietveld-fit (red) and difference curve (gray). Insert: crystal structure of [(Li$_{0.8}$Fe$_{0.2}$)OH]Fe(S$_{1-x}$Se$_x$).

The tetragonal structure consists of *anti*-PbO type layers of lithium-iron-hydroxide alternating with FeS layers. X-ray single crystal analysis confirms the structure. Crystallographic data as well as X-ray powder patterns for $x > 0$ are compiled in the electronic supplementary information (ESI)†. The compositions of all compounds were confirmed combining energy dispersive X-ray spectroscopy (EDX) measurements, inductively coupled plasma (ICP) analysis and elementary analysis. Remarkably, the composition of the $(Li_{1-x}Fe_x)OH$ layer in the new compounds is the same as in $[(Li_{1-x}Fe_x)OH](Fe_{1-y}Li_y)Se$[7-9] which means that the same charge transfer of 0.2 electrons takes place in the sulphide. An open issue regarding the crystal structure is the large $U_{33}$ component of the thermal displacement ellipsoid at the Fe/Li mixed site in all compounds. This was also observed by *Sun* et al.,[9] and may be interpreted as a split position with Li shifted off the centre of the oxygen tetrahedra along [001]. Contrary to $[(Li_{1-x}Fe_x)OH](Fe_{1-y}Li_y)Se$ where the presence of Li or alternatively iron vacancies in the FeSe layer is discussed,[7] refinements of X-ray single crystal diffraction data gives no indication of a Fe/Li mixed site or iron vacancies at this position on the sulphur doped compounds. *Sun* et al. suggested that the lattice parameter *a* decreases with increasing amount of Fe vacancies in the FeSe layer.[9] The lattice parameter *a* of $[(Li_{0.8}Fe_{0.2})OH]FeS$ is 370 pm, distinctly smaller compared to the selenides with $a = 378 - 382$ pm[9], thus a Fe/Li mixed site or iron vacancies in the FeS layer is unlikely.

The lattice parameters as well as the unit cell volumes of $[(Li_{0.8}Fe_{0.2})OH]Fe(S_{1-x}Se_x)$ increase linearly with the doping level *x* as shown in Fig. 2.

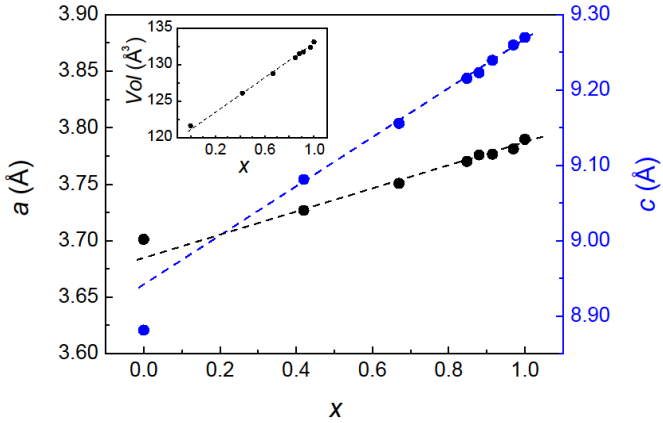

**Fig. 2** Lattice parameters *a* (black) and *c* (blue) of $[(Li_{0.8}Fe_{0.2})OH]Fe(S_{1-x}Se_x)$. Insert: unit cell volume. Dashed lines are guides to the eye.

The linear behaviour of *a* and *c* within the whole doping range $0 \leq x < 1$ indicates homogeneous doping of sulphur, however, the pure sulphide slightly deviates from linearity. The shrinking of the unit cell due to the smaller ionic radius of sulphur is also known from *anti*-PbO type $Fe(Se_{1-z}S_z)$ ($z = 0-0.5$) with a possible solubility limit of $z \approx 0.3$.[13] The critical temperature of $Fe(Se_{1-z}S_z)$ increases up to 15.5 K for $x \leq 0.2$ due to chemical pressure.[13] $T_c$ decreases again at $x \geq 0.3$, thus it remains much smaller under chemical than under physical pressure (36 K).[14] In contrast, sulfur-doping of $[(Li_{0.8}Fe_{0.2})OH]Fe(S_{1-x}Se_x)$ continuously decreases $T_c$ linearly, until superconductivity is completely suppressed in the pure sulphide, as seen from the *dc* electrical transport measurements in Fig. 3.

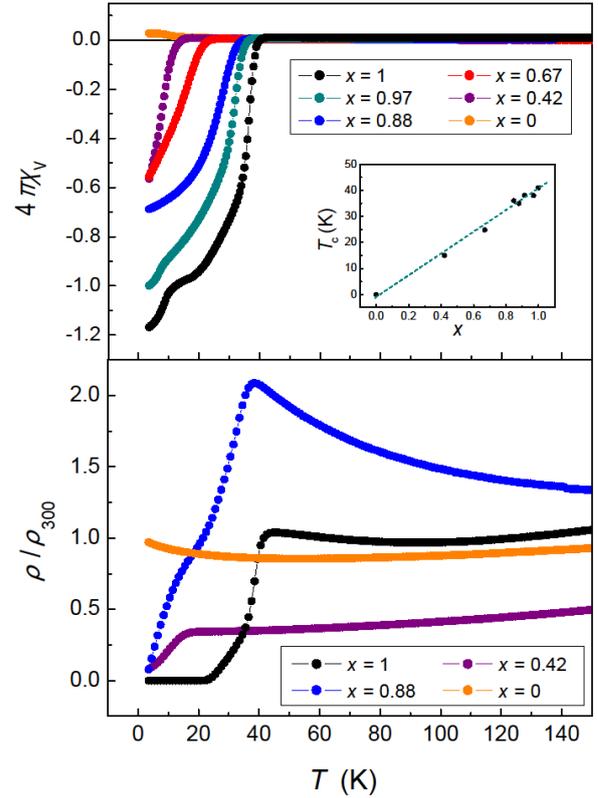

**Fig. 3** Top: *ac*-susceptibility of $[(Li_{0.8}Fe_{0.2})OH]Fe(S_{1-x}Se_x)$. Insert: development of $T_c$ with *x*. Bottom: dc resistivity for $x = 0$ (orange), 0.42 (magenta), 0.88 (blue) and 1 (black).

The *dc*-resistivity of the pure selenide compound is weakly temperature dependent until it drops abruptly at 43 K (lower panel in Fig. 3). For $x = 0.88$ the resistivity drop is shifted to 37 K, and a shoulder appears at about 20 K, which is most probably due to the magnetic ordering of the Fe moments in the $(Li_{0.8}Fe_{0.2})OH$ layer (see below). The relatively large increase of resistivity above the superconducting transition is due to grain boundary effects, because cold pressed pellets have to be used owing to the temperature sensibility of the compounds. For $x = 0.42$ a distinct drop in resistivity is discernible at about 15 K, which is in good agreement with magnetic susceptibility measurements. As in this case the superconducting transition temperature coincides with the temperature range where the ferromagnetic ordering arises, the decrease in resistivity is rather broad. A tiny residual resistivity is observed, caused by grain boundary effects of the cold pressed pellets. Undoped $[(Li_{0.8}Fe_{0.2})OH]FeS$ shows no sign of a superconducting transition, which is in line with the magnetic susceptibility measurements. However, a slight increase in resistivity can be observed for low temperatures, which might be again due to the gradual magnetic ordering in the $(Li_{0.8}Fe_{0.2})OH$ layer.

The enormous increase of $T_c$ in $[(Li_{0.8}Fe_{0.2})OH](Fe_{1-y}Li_y)Se$ (43 K) in comparison to β-FeSe (8 K) can be explained by electron doping from the hydroxide to the selenide layer.[7] By substituting Se by S this electron doping doesn't change as Se and S ions have the same valence and the composition of the hydroxide layer remains constant. Though, the smaller S atoms lead to a chemical pressure effect, which additionally influences superconductivity. Contrary to $Fe(Se_{1-z}S_z)$ where chemi-cal pressure enhances superconductivity, we observe a decrease of $T_c$ with increasing chemical pressure. Apparently in our case the geometry of the tetrahedral Fe(S$_{1-x}$Se$_x$) layer is not further optimized. With increasing amount of S the unit cell volume and the Fe-Fe as well as the Fe-(Se,S) distances shrink. Contemplating the $Ch$-Fe-$Ch$ bond angles of the Fe$Ch_4$ tetrahedra, a flattening of the Fe(S$_{1-x}$Se$_x$) layers with increasing sulfur doping is observed (for the respective diagrams see ESI). A definitive clue which parameter is crucial with respect to $T_c$ cannot be given at this point. However, an enlargement of the unit cell with the respective opposite evolutions in geometry of the FeSe layer by substituting Se by Te seems quite promising in view of an enhancement of $T_c$.

The possible coexistence of superconductivity and ferromagnetism in the series [(Li$_{0.8}$Fe$_{0.2}$)OH]Fe(S$_{1-x}$Se$_x$) is of particular interest. In [(Li$_{0.8}$Fe$_{0.2}$)OH](Fe$_{1-y}$Li$_y$)Se magnetic ordering emerges from the iron ions in the hydroxide layer at about 10 K, well below the superconducting transition temperature at 43 K. In contrast to the suppression of the critical temperature with increasing sulphur doping, ferromagnetism persists over the whole substitution range. Figure 4 shows the magnetic susceptibility of [(Li$_{0.8}$Fe$_{0.2}$)OH]FeS (black) and doped samples.

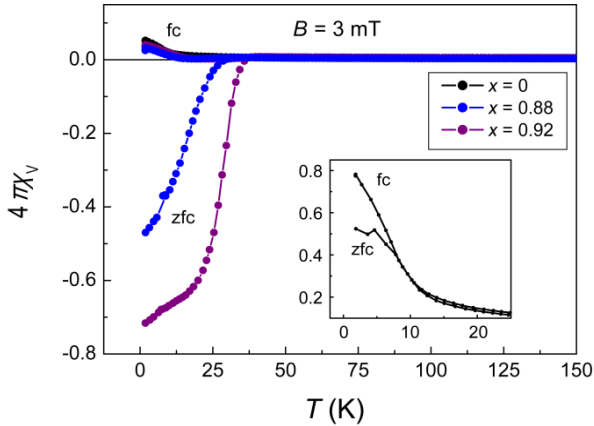

**Fig. 4** Magnetic dc-susceptibility of [(Li$_{0.8}$Fe$_{0.2}$)OH]Fe(S$_{1-x}$Se$_x$) for $x = 0$ (black), 0.88 (blue) and 0.92 (magenta). Insert: magnification of the low-temperature part for $x = 0$.

Selenium rich compounds show a strong diamagnetic signal in a 3 mT field analogous to [(Li$_{0.8}$Fe$_{0.2}$)OH](Fe$_{1-y}$Li$_y$)Se. However, this holds true only for the zero-field cooled mode (zfc, Figure 4) where the shielding effect is strong. After field-cooling (fc, Figure 4) the susceptibility becomes merely slightly negative below $T_c$ owing to the Meissner-Ochsenfeld effect before increasing to positive values at lower temperatures. This behaviour is also known from [(Li$_{0.8}$Fe$_{0.2}$)OH](Fe$_{1-y}$Li$_y$)Se and a result of the coexistence of ferromagnetism and superconductivity.[7] The susceptibility of undoped [(Li$_{0.8}$Fe$_{0.2}$)OH]FeS is throughout positive as superconductivity is completely suppressed. Nevertheless, for low temperatures we also observe a different signal in zfc and fc mode, respectively (see insert in Fig. 4). This splitting is typical for ferromagnetic ordering and caused by different domain formations in fc and zfc modes. Below $T_{fm} \approx 10$ K, the magnetic moments order spontaneously leading to an increase in magnetic susceptibility. In zfc mode, the domains are randomly distributed. Switching on the external field the domains tend to orientate along the field which is only partially accomplished. As a result the signal is lower compared to fc mode where the domains can align in the field during the cooling cycle.

The inverse susceptibility of [(Li$_{0.8}$Fe$_{0.2}$)OH]FeS at 2 T obeys the Curie-Weiss law with an effective magnetic moment of 4.98(1) $\mu_B$ (see ESI). This value is in good accordance with the theoretically expected one of 4.9 $\mu_B$ for Fe$^{2+}$ contrary to the one of Fe$^{3+}$ (5.9 $\mu_B$).[15] Thus, the situation of the iron ions in the hydroxide layer is unchanged. It seems obvious that the electron transfer to the Fe(S$_{1-x}$Se$_x$) layer and magnetic ordering in the (Li$_{0.8}$Fe$_{0.2}$)OH layer persist in [(Li$_{0.8}$Fe$_{0.2}$)OH]Fe(S$_{1-x}$Se$_x$) in the whole substitution range.

The interplay of magnetism and superconductivity is further confirmed by magnetization measurements (Fig. 5).

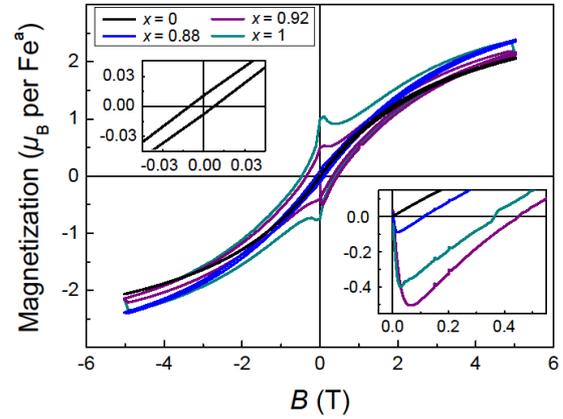

**Fig. 5** Isothermal magnetization at 1.8 K of [(Li$_{0.8}$Fe$_{0.2}$)OH]Fe$^a$(S$_{1-x}$Se$_x$) for $x = 0$ (black), 0.88 (blue), 0.92 (magenta) and 1 (dark cyan). Insert: magnification of the low-field part showing the hysteresis for $x = 0$ (left insert) and the initial curves (right insert).

The ferromagnetic hysteresis of [(Li$_{0.8}$Fe$_{0.2}$)OH]Fe(S$_{1-x}$Se$_x$) with $x > 0$ is superimposed by the magnetization known for hard type-II superconductors.[7] The initial curves prove superconductivity for the Se containing compounds, which is in line with susceptibility measurements. Decreasing the amount of Se, the superconducting hysteresis continuously diminishes. As expected from susceptibility measurements, [(Li$_{0.8}$Fe$_{0.2}$)OH]FeS shows only the ferromagnetic ordering with a very narrow hysteresis typical for a soft ferromagnet. We suppose that the reason for this is the dilution of the magnetic iron ions in the hydroxide layer leading to small coupling.

## Conclusions

[(Li$_{0.8}$Fe$_{0.2}$)OH]FeS and the series [(Li$_{0.8}$Fe$_{0.2}$)OH]Fe(S$_{1-x}$Se$_x$) were synthesized by hydrothermal methods and characterized by X-ray single crystal and powder diffraction, EDX and chemical analysis. Selenium-rich compounds show coexistence of magnetic ordering with superconductivity as known from the pure selenium compound. Sulphur doping decreases the critical temperature through chemical pressure until superconductivity is completely absent in [(Li$_{0.8}$Fe$_{0.2}$)OH]FeS, while ferromagnetism persists in the [(Li$_{0.8}$Fe$_{0.2}$)OH] layers. The Li:Fe ratio in the hydroxide layer and thus the charge transfer of 0.2 electrons from the hydroxide to the iron chalcogenide layers remains unchanged in [(Li$_{0.8}$Fe$_{0.2}$)OH]Fe(S$_{1-x}$Se$_x$), which indicates that the chemical pressure effect of the smaller sulphide ions impedes superconductivity in [(Li$_{0.8}$Fe$_{0.2}$)OH]FeS.

## Acknowledgement

This work was financially supported by the German Research Foundation (DFG) within the priority program SPP1458 and by the FP7 European project SUPER-IRON (Grant no. 283204).

## Notes and references

*Department Chemie, Ludwig-Maximilians-Universität München, Butenandtstrasse 5-13 (Haus D), 81377 München (Germany);*
*E-Mail: johrendt@lmu.de*

† Electronic Supplementary Information (ESI) available: [Table of crystallographic data, X-ray powder patterns of $[(Li_{0.8}Fe_{0.2})OH]Fe(S_{1-x}Se_x)]$ ($0 \leq x \leq 1$), plots showing the evolution of Fe-$Ch$ distances and $Ch$-Fe-$Ch$ bond angles with $x$, and Curie-Weiss fit for $[(Li_{0.8}Fe_{0.2})OH]FeS$. See DOI: 10.1039/c000000x/

‡ Materials: Fe powder (Chempur, 99.9 %), Selenourea (Alfa Aesar, 99 %), Thiourea (Grüssing, 99 %), LiOH (Fisher Scientific).

X-ray powder diffraction was carried out using a Huber G670 diffractometer with Cu-K$\alpha_1$ radiation ($\lambda = 154.05$ pm) and Ge-111 monochromator. Structural parameters were obtained by Rietveld refinement using the software package TOPAS.[16] Single-crystal analysis was performed on a bruker D8-Quest diffractometer (Mo-K$\alpha_1$, $\lambda = 71.069$ pm, graphite monochromator). The structure was solved and refined with the Jana2006 program package.[17] Chemical compositions were additionally determined by energy-dispersive X-ray analysis (EDX) as well as by chemical methods using ICP-AAS and elemental analysis. Magnetic properties were examined with a Quantum Design MPMS-XL5 SQUID magnetometer, whereas superconductivity was examined in ac-susceptibility measurements. Temperature-dependent resistivity measurements were carried out on cold pressed pellets using a standard four-probe method.